\documentclass[aps,twocolumn,amsmath,floatfix,superscriptaddress]{revtex4-1}
\usepackage{lineno,isotope,mathrsfs}
\modulolinenumbers[5]
\usepackage{amsmath,bbold,amssymb,epsfig,bm,mathrsfs,feynmp}
\usepackage{color,slashed,exscale,multirow,soul}
\usepackage{times,txfonts,xcolor,graphicx,tikz}
\usepackage[normalem]{ulem}
\usepackage[breaklinks,colorlinks,citecolor=blue]{hyperref}
\definecolor{red}{rgb}{0.8,0,0}
\definecolor{violet}{rgb}{0.4,0,0.4}
\definecolor{green}{rgb}{0,0.5,0.0}
\definecolor{navy}{rgb}{0.0,0.0,0.6}
\definecolor{orange}{rgb}{0.8,0.2,0.0}

\usepackage[normalem]{ulem}  

\newcommand{\bea}{\begin{eqnarray}}
\newcommand{\eea}{\end{eqnarray}}
\newcommand{\ep}{\varepsilon}
\newcommand{\MR}{$M$-$R\ $}

\newcommand{\Lsym}{\ensuremath{L_{\text{sym}}}}
\newcommand{\Qsat}{\ensuremath{Q_{\text{sat}}}}
\newcommand{\rhotran}{\ensuremath{\rho_{\text{tran}}}}

\begin{document}
\title{Ultracompact hybrid stars consistent with multimessenger astrophysics}
\author{Jia Jie Li}
\affiliation{School of Physical Science and Technology, 
Southwest University, Chongqing 400715, China}
\author{Armen Sedrakian}
\affiliation{Frankfurt Institute for Advanced Studies,
D-60438 Frankfurt am Main, Germany}
\affiliation{Institute of Theoretical Physics,
University of Wroc\l{}aw, 50-204 Wroc\l{}aw, Poland}
\author{Mark Alford}
\affiliation{Department of Physics, Washington University,
St.~Louis, Missouri 63130, USA}
\begin{abstract}
  In this work, we consider the consequences of phase transition in
  dense QCD on the properties of compact stars and implications for
  the observational program in gravitational wave and X-ray
  astrophysics.  The key underlying assumption of our modeling is a
  strong first-order phase transition past the point where the
  hadronic branch of compact stars reaches the two-solar mass limit.
  Our analysis predicts ultracompact stars with very small radii - in
  the range of 6-9 km - living on compact star sequences that are
  entirely consistent with the current multimessenger data. We show
  that sequences featuring two-solar mass hadronic stars consistent
  with radio-pulsar observations are also consistent with the inferences 
  of large radii for massive neutron stars by NICER X-ray observations 
  of neutron stars and the small radii predicted by gravitational 
  waves analysis of the binary neutron star inspiral event GW170817  for our models that feature a strong first-order QCD phase transition.

\end{abstract}

\date{\today}
\maketitle
\section{Introduction}
There has been significant progress in the search for the ultimate
state of extremely dense (by an order of magnitude larger than in the
ordinary nuclei) matter over the last decade due to new astrophysical
observations of compact stars. These include the GW170817 event
involving a merger of two compact stars which heralded the beginning
of the multimessenger era of exploration of compact
stars~\cite{LIGO_Virgo2017,LIGO_Virgo2018,LIGO_Virgo2019}, the X-ray 
observation of nearby neutron stars by the NICER
instrument~\cite{NICER2019a,NICER2021a,NICER2019b,NICER2021b}, and
measurement of massive radiopulsars in binaries with white
dwarfs~\cite{NANOGrav2019}.

The description of ultra-dense matter should ultimately be based on
first-principles QCD which is currently out of reach in the
density-temperature range relevant for compact stars. Thus, a
combination of phenomenological models with the available
experimental/astrophysical data is the best option to model the
properties of dense, strongly-interacting states of matter, for
reviews see Refs.~\citep{Alford2008,Anglani2014}. A likely outcome of
the compression of the nucleonic matter is a phase transition to
liberated quark phase, as envisioned already several decades
ago~\citep{Ivanenko1965,Itoh1970,Collins1975} and extensively studied
over the past decades. It has been realized that a first-order phase
transition between the hadronic and quark phase may lead to a new
branch of stable hybrid stars---stars featuring a dense quark core
enveloped by nucleonic matter.  One interesting feature of hybrid CS is the existence of {\em twin
  configurations}: two stars with different radii but the same masses,
the larger one being purely nucleonic and the more compact one being a
hybrid star. These hybrid stars comprise the {\it third family of 
compact stars} after white dwarfs and neutron
stars~\cite{Gerlach1968,Bowers1977,Kaempfer1981,Glendenning2000,Schertler2000,Benic2015,Alford2017,Alvarez-Castillo2019,Han:2018mtj,Otto2020,Lijj2021,Li:2019fqe,Christian2021b}.
Below we will use the specific constant speed of sound
parametrization of the quark phase~\cite{Zdunik2013,Alford2013}, but
qualitatively similar results are expected for alternative models,
for example, multpolytropic EoS~\cite{Alvarez-Castillo2019}. Indeed,
modeling based on the CSS and multipolytropic EoS leads to
qualitatively similar results, see Ref.~\cite{Paschalidis:2017qmb}.  

Here we address the structure of hybrid stars which are consistent with current
astrophysical observations and show that a strong first-order phase transition may lead to the emergence of {\it hybrid stars with
extremely small radii and a very narrow range of masses for any 
given parameter set. However, the masses of these objects can be 
varied broadly by changing the parameters of the equation of state
(EoS) while keeping the resultant stellar sequences consistent with 
current multimessenger astrophysical data.}  This sheds new light on
the ways the current astrophysical constraints can be interpreted. For
example, our analysis below confirms that, as previously noted
\cite{Lijj2021,Christian2021b}, the masses and radii inferred from
NICER observations can be interpreted as originating from the
nucleonic branch of mass-twins, while the hybrid branch contains stars
whose tidal deformabilities (hereafter TD) are consistent with the
GW170817 observation. Another implication of our study is a shift in
the paradigm stating that a small $R\simeq 6$-9~km radius objects, if
discovered, would be strange stars (for a review see
Ref.~\cite{Weber2005}). As we show below, ultracompact objects
arising from a strong first-order phase transition can have radii
covering the range where so far only strange stars (objects arising
within the Witten-Bodmer conjecture about the ground state of absolute
stable state of matter~\cite{Witten1984,Bodmer1971}) were
predicted. Previously, Drago et al.~\cite{Drago2013} addressed the
compatibility of large masses and small radii, in terms of two
separate families of compact stars. Their scenario invokes very
compact hadronic stars, whose EoS is soft, and strange stars whose EoS
is stiff.  The strange star hypothesis by itself is compatible 
with the NICER data~\cite{Horvath2021IJMPD} and deformabilities 
inferred from GW170817~\cite{Lourenco2021PhRvD}, but has been the
 only one that accounted for very small radii in the range of
6-9~km. As we show below, large masses and small radii are
possible to accommodate within a single-family scenario, which needs
to be incorporated in the analysis of the current astrophysics data.

\section{Constructing the equation of state}
To model the EoS of low-density hadronic matter we use the covariant density functional (CDF) approach based on the  Lagrangian of stellar matter with baryonic degrees of freedom 
$
\mathscr{L} = \mathscr{L}_b + \mathscr{L}_m + \mathscr{L}_l + \mathscr{L}_{\rm em},
$
where the baryon Lagrangian is given by
\begin{eqnarray}
\label{eq:lagrangian}
\mathscr{L}_b  = \sum_b\bar\psi_b\Big[\gamma^\mu
\big(i\partial_\mu-g_{\omega b}\omega_\mu
      -g_{\rho b}{\mathbf{\tau}}\cdot\mathbf{\rho}_\mu\big)
      -\big(m_b - g_{\sigma b}\sigma\big)\Big]\psi_b , 
      \nonumber \\
\end{eqnarray}
with the $b$-sum running over the $J^P = \frac{1}{2}^+$ baryon octet in general, but we restrict the discussion to nucleons only; $\psi_b$ are the nucleonic Dirac fields with masses $m_b$, and $\sigma,\omega_\mu$, and $\mathbf{\rho}_\mu$ are the mesonic fields which mediate the interaction among baryon fields. The remaining pieces of the Lagrangian correspond to the mesonic, leptonic, and electromagnetic contributions, respectively. These are standard and are given, e.g., in Ref.~\cite{Sedrakian2021}. The density-dependent nucleon-meson couplings $g_{mb}$ are fixed at saturation density at the values prescribed by the DDME2 parametrization~\citep{Lalazissis2005}. Their density dependence differs from that parametrization and is varied to match  
the resulting EoS with the phenomenological expansion of the 
energy density of nuclear matter~\cite{Margueron2018,Lijj2019b,Libaoan2021}
\begin{eqnarray}
\label{eq:Taylor_expansion}
E(\chi, \delta) & \simeq & E_{\rm{sat}} + \frac{1}{2!}K_{\rm{sat}}\chi^2
                             + \frac{1}{3!}Q_{\rm{sat}}\chi^3 \nonumber \\ [0.5ex]
                        &  & +\,E_{\rm{sym}}\delta^2 + L_{\rm{sym}}\delta^2\chi
                             + {\mathcal O}(\chi^4, \chi^2\delta^2),
\end{eqnarray}
where $\chi\equiv(\rho-\rho_{\rm{sat}})/3\rho_{\rm{sat}}$,
$\rho_{\rm sat}$ is the saturation density, and
$\delta = (\rho_{\rm n}-\rho_{\rm p})/\rho$ is the isospin asymmetry with
$\rho_{\rm n(p)}$ being the neutron(proton) number densities.
The coefficients in this double expansion are referred to commonly as 
the {\it incompressibility} $K_{\rm{sat}}$, {\it skewness} $Q_{\rm{sat}}$,
{\it symmetry energy} $E_{\rm{sym}}$, and its {\it slope}
$L_{\rm{sym}}$. The mapping between the CDF and the phenomenological
expansion \eqref{eq:Taylor_expansion} allows us to express the gross
properties of compact stars in terms of physically transparent
quantities.

\begin{table}[b]
\centering
\caption{
The meson masses and the meson-nucleon coupling constants  at the nuclear saturation density (column I), the constants determining the density 
dependence of the meson-nucleon couplings for
$Q_{\rm {sat}} = 300$ (column II) and 900\,MeV (column III). The last two rows of columns II and III list the couplings that produce 
$L_{\rm {sym}} = 45$ (left entry) and 105\,MeV (right entry).}
\setlength{\tabcolsep}{8pt}
\label{tab:Parameters1}
\centering
\begin{tabular}{ccccc}
\hline\hline
           &          &$Q_{\rm{sat}}$&  300          &   900    \\
\hline    &    I       &            &
          II           &  III       \\
\hline
$m_\sigma$ & 550.1238 & $a_\sigma$   & 1.3690        & 1.4730        \\
$m_\omega$ & 783.0000 & $b_\sigma$   & 0.8555        & 1.9201        \\
$m_\rho$   & 763.0000 & $c_\sigma$   & 1.3353        & 3.0965        \\
$g_\sigma$ &  10.5396 & $d_\sigma$   & 0.4996        & 0.3281        \\
$g_\omega$ &  13.0189 & $a_\omega$   & 1.3752        & 1.4571        \\
           &          & $b_\omega$   & 0.7205        & 1.6107        \\
           &          & $c_\omega$   & 1.1493        & 2.5947        \\
           &          & $d_\omega$   & 0.5385        & 0.3584        \\
           &          & $g_\rho$     & 3.3379/4.2193 & 3.3253/4.2111 \\
           &          & $a_\rho$     & 0.6442/0.0506 & 0.6552/0.0569 \\
\hline\hline
\end{tabular}
\end{table}
In this work, we study a family of representative nucleonic EoS
obtained by varying $\Lsym$ and $Q_{\rm sat}$ at fixed values of
$E_{\rm{sym}}(\rho_c) = 27.09$\,MeV ($\rho_c = 0.11$\,fm$^{-3}$) and 
$K_{\rm{sat}} = 251.15$\,MeV~~\cite{Lijj2019b}. The parameter 
$Q_{\rm sat}$ controls the high-density behavior of the EoS,
and thus, the maximum mass of a static nucleonic CS, whereas
$L_{\rm sym}$ controls the intermediate-density EoS and is strongly 
correlated with the radius of the nucleonic stars. We consider 
values of the slope parameter in the range 
$45 \le L_{\rm sym} \le 105$~MeV. The upper limit of $L_{\rm sym}$ 
is the central value of the PREX-II
measurement interpretation by Refs.~\citep{PREX-II2021,Reed2021}; the
lower bound corresponds to the one derived from the analysis of the
same experimental data in Ref.~\cite{Reinhard2021}.  The values for
skewness are less constrained and we use values $Q_{\rm sat}= 300$ and
900~MeV which predict maximal masses of nucleonic sequences within the
mass range $2.30 \le M_{\rm max}/M_{\odot} \le 2.55$.  This mass range
allows for the nucleonic CS branch to account for the measured lower bound
on the maximum mass $M/M_{\odot}= 2.08\pm 0.07$~\cite{Fonseca2021}.
The pairs of $\Qsat$ and $\Lsym$ values bracket the range of accepted values of these parameters. Our analysis shows that qualitatively similar results are obtained when using other pairs of 
these parameters that are drawn from the bracketed range.  The parameters for four nucleonic EoS models are listed  in Table \ref{tab:Parameters1}.

The EoS of the quark phase is modeled by a constant sound speed (CSS)
parametrization~\cite{Zdunik2013,Alford2013} which offers a {\it synthetic} 
model motivated by microscopic computations~\citep{Alford2017}, i.e.,
\bea\label{eq:EoS}
p(\ep) =\left\{
\begin{array}{ll}
p_{\rm tran}, \,\, & \ep_{\rm tran} < \ep < \ep_{\rm tran}\!+\!\Delta\ep, \\[0.5ex]
p_{\rm tran} + s\,\bigl[\ep-(\ep_{\rm tran}\!+\!\Delta\ep)\bigr],
  \,\, & \ep_{\rm tran}\!+\!\Delta\ep < \ep, 
\end{array}
\right.
\eea
where $p_{\rm tran}$ and $\ep_{\rm tran}$ are the pressure and energy
density at which the transition from hadronic (hereafter H) to
high-density quark phase (hereafter Q) takes place; $s$ is the squared
speed of sound (in natural units) in the quark matter phase and
$\Delta\ep$ is the discontinuity in energy density; note that there is
no state with energy between $\ep_{\rm tran}$ and
$\ep_{\rm tran}\!+\!\Delta\ep$. This parametrization agrees with the
predictions of computations based on the Nambu--Jona-Lasinio (NJL)
model supplemented by vector
repulsion~\citep{Blaschke2010,Bonanno2012,Blaschke2013,Pfaff2022}. In
constructing our models we will use the Maxwell construction to match
our nucleonic EoS to the quark one~\eqref{eq:EoS}.

\section{The mass-radius diagram and tidal deformabilities}
\begin{figure}[tb]
\centering
\ifpdf
\includegraphics[width = 0.45\textwidth]{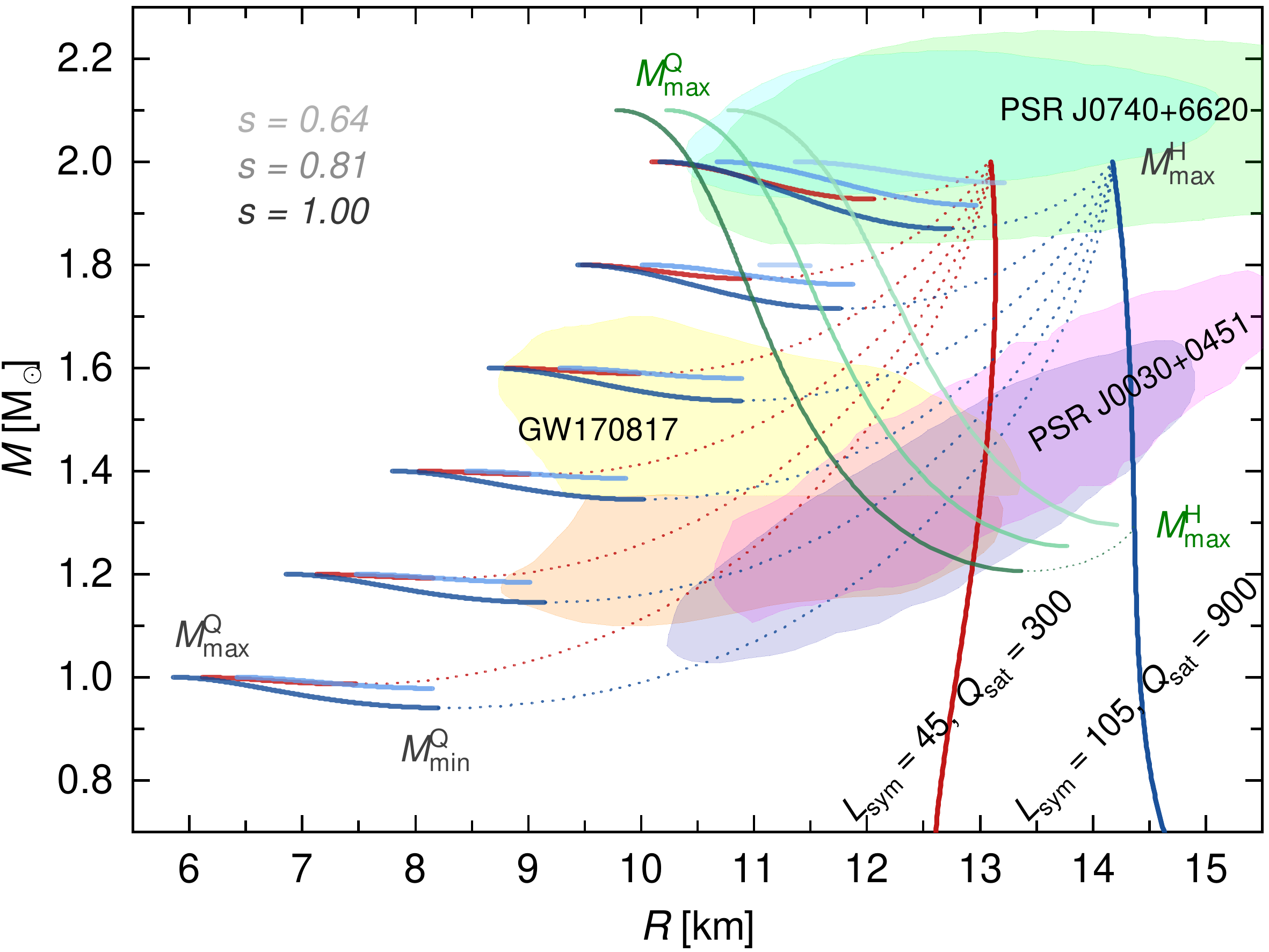}
\else
\includegraphics[width = 0.45\textwidth]{MR1.eps}
\fi
\caption{ \MR relation for hybrid EoS models with a phase transition
  at high density for stiff-stiff nucleonic EoS with $\Lsym = 105,\,\Qsat = 900$~MeV, and soft-soft one with $\Lsym = 45,\,\Qsat = 300$~MeV.
  The sequences of stars are constructed from a nucleonic model by
  fixing $M^{\rm H}_{\rm max}/M_{\odot} = 2.0$ on the nucleonic
  branch, while varying the $M^{\rm Q}_{\rm max}/M_{\odot}$ in the
  range 1.0-2.0 and the speed of sound squared $s$. For comparison we
  also, show the sequences in the case of low-maximum hadronic mass
  $M^{\rm H}_{\rm max}/M_{\odot} = 1.3$, studied previously in Refs.~\cite{Alvarez-Castillo2019,Lijj2021}, for the same speeds of sound. Note that GW170817 
  ellipses assume a hadronic EoS and constrain only the hadronic 
  branches of the stellar sequences.}
\label{fig:illust}
\end{figure}

The three quark matter parameters ($p_{\rm tran}$, $\Delta\ep$,
$s$) fully determine the mass-radius (hereafter $M$-$R$) curves 
for hybrid stars for any given nucleonic EoS. It is convenient for 
further discussion to denote maximum/minimum masses for the branches by
$M^{\rm H}_{\rm max}$, $M^{\rm Q}_{\rm max}$, and
$M^{\rm Q}_{\rm min}$.  See Fig.~\ref{fig:illust} for an illustration of
these parameters and the features of the \MR diagram, in particular 
the maxima and minima that may arise. Note that the branch extending 
up to $M^{\rm Q}_{\rm max}$ corresponds to the third family.

Figure~\ref{fig:illust} shows the scenario under consideration in
which the phase transition occurs at a high enough density,
$\rhotran/\rho_{\rm sat}\gtrsim 3.0$, so that the heaviest star on the
hadronic branch has a mass of $2\,M_{\odot}$.  In this case, the
hybrid branches lie at lower masses and are nearly flat. In the same
figure, we contrast this scenario with the one studied in the light of
multimessenger data in Refs.~\citep{Alvarez-Castillo2019,Lijj2021}, in
which the phase transition occurs at a lower density,
$\rhotran/\rho_{\rm sat} \lesssim 2.0$, causing the hadronic branch to
end at a lower mass, $M^{\rm H}_{\rm min}=1.3\,M_{\odot}$, and the
hybrid branch extends up to $M^{\rm Q}_{\rm max} > 2\,M_{\odot}$.  Our
sequences can be confronted with current astrophysical observational
constraints shown also in Fig.~\ref{fig:illust} which include: (a)~the
ellipses indicating the regions of \MR diagram compatible with the
analysis of NICER observations of PSR~J0030+0451 and
J0740+6620~\citep{NICER2019a,NICER2021a,NICER2019b,NICER2021b};
(b)~the regions of \MR diagram that are compatible with the parameters
of the two compact stars that merged in the gravitational wave event
GW170817~\citep{LIGO_Virgo2018}. For both observations, the ellipses
show the 90\% credible intervals (CIs). Note that ellipses
referring to GW170817 have been obtained under the assumption of a
hadronic star, therefore they are relevant, strictly speaking, only
for constraining the hadronic branches on the \MR diagram.
Figure~\ref{fig:illust} shows, in addition, the sensitivity of results
to varying the quark matter speed of sound squared $s$. We see that
choosing the maximum value $s=1.0$ yields the widest range of masses
on the hybrid branch and hence for twin stars.

\begin{figure}[tb]
\centering
\ifpdf
\includegraphics[width = 0.45\textwidth]{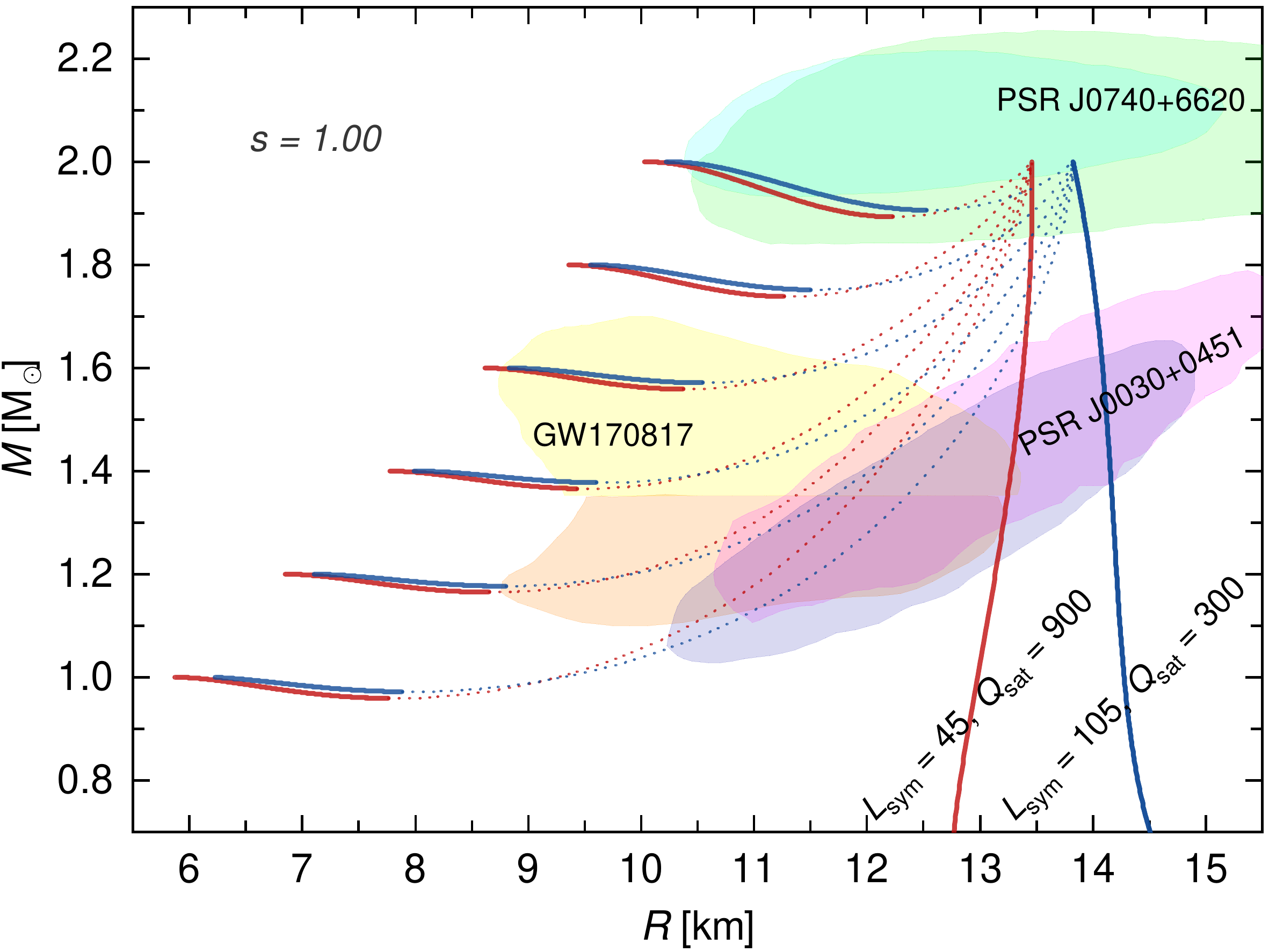}
\else
\includegraphics[width = 0.45\textwidth]{MR2.eps}
\fi
\caption{ \MR relation for hybrid EoS models featuring high density
phase transition with two examples of nucleonic EoS, specifically the
soft-stiff one with $\Lsym = 45, \Qsat = 900$~MeV and stiff-soft one
$\Lsym = 105, \Qsat = 300$ MeV. The value of sound speed is fixed at
$s = 1.0$.}
\label{fig:MR_twi}
\end{figure}

In Fig.~\ref{fig:MR_twi} we fix the maximum mass of the nucleonic
branch $M^{\rm H}_{\rm max} = 2.0\,M_{\odot}$ and squared sound of
speed $s = 1.0$ and then vary $M^{\rm Q}_{\rm max}/M_{\odot}$ from 1.0
to 2.0 to show the range covered in the \MR-diagram by these types of
sequences.

To assess the range of variations in the sequences arising from the
uncertainties in the {\it nucleonic sector} we consider nucleonic EoS
with $L_{\rm sym}$ taking values $45, 105$\,MeV and $Q_{\rm sat}$
taking values $300, 900$\,MeV.  The value of $L_{\rm sym}$ controls
the intermediate-density and $Q_{\rm sat}$ the high-density behavior
on the nucleonic branch, see Figs.~\ref{fig:illust} and
\ref{fig:MR_twi}.  So in Fig.~\ref{fig:MR_twi} we have
$(L_{\rm sym},Q_{\rm sat})=(105, 300)$\,MeV corresponding to a
``stiff-soft'' (intermediate-density stiff and high-density soft)
hadronic EoS and $(L_{\rm sym},Q_{\rm sat})=(45, 900)$\,MeV for
the inverse ``soft-stiff'' EoS. 

\begin{figure}[tb]
\centering
\ifpdf
\includegraphics[width = 0.45\textwidth]{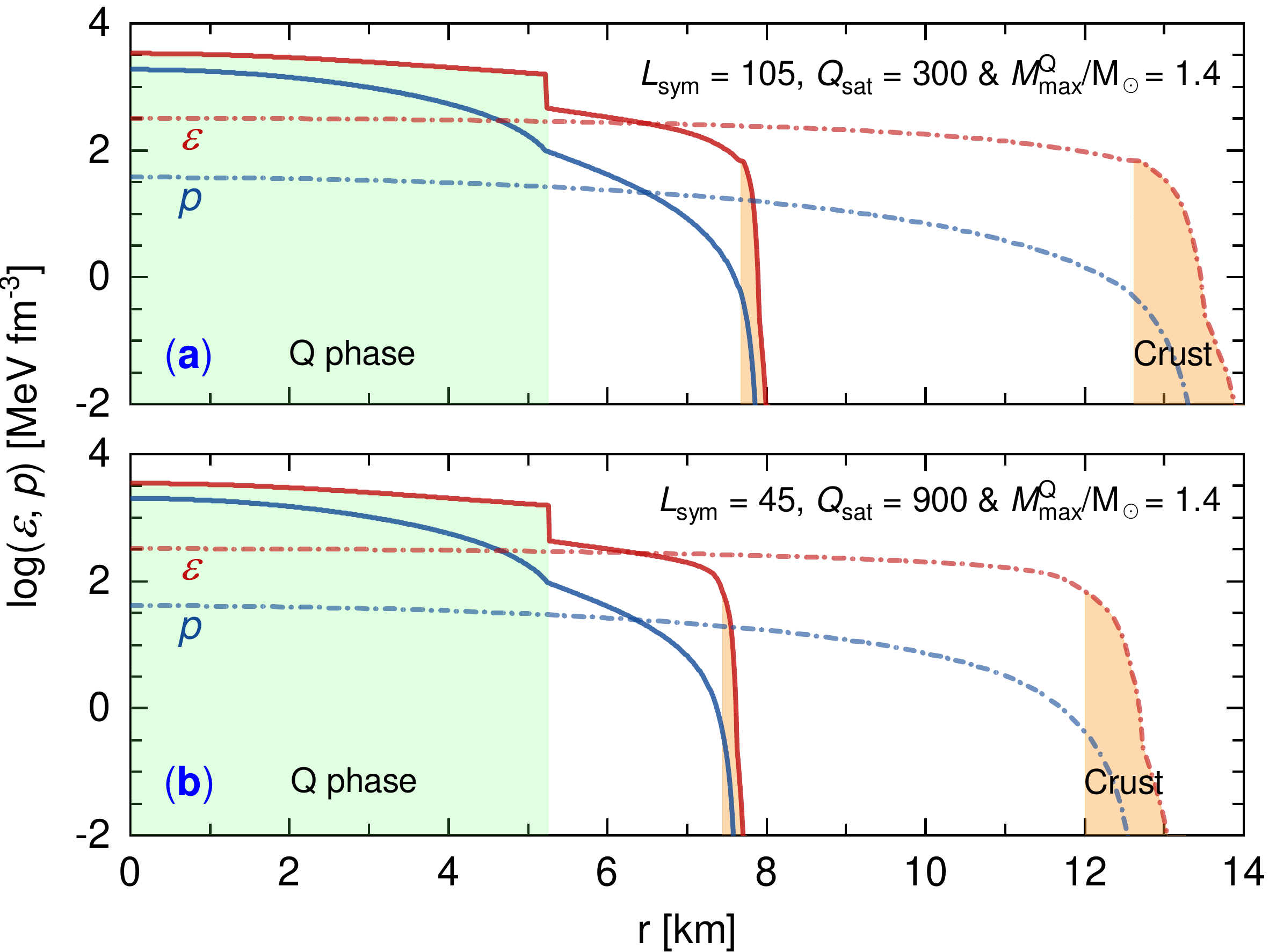}
\else
\includegraphics[width = 0.45\textwidth]{EP_R.eps}
\fi
\caption{Radial profiles of energy density and pressure (on
  a logarithmic scale) for twin stars with masses
  $M = M^Q_{\rm max} =1.40\,M_{\odot}$. The results for purely
  nucleonic stars are shown by dotted, and those for hybrid stars - by
  solid lines.  Two nucleonic EoS have been used: a soft-stiff one
  with $\Lsym = 45, \Qsat = 900$~MeV and a stiff-soft one with
  $\Lsym = 105, \Qsat = 300$ MeV.  The quark core and crust regions
  are shaded for clarity.  }
\label{fig:EP_R}
\end{figure}

Figure~\ref{fig:EP_R} shows the internal structure of twin stars with
masses $M = 1.40\,M_{\odot}$ by plotting the energy density and
pressure as a function of distance from the center for purely
nucleonic (dotted lines) and hybrid (solid lines) stars. The highly
compact hybrid configurations have an appearance that is similar to
frequently studied less compact hybrid stars. As seen in
Fig.~\ref{fig:EP_R} there is a moderate-size quark core of about 5~km,
a nucleonic layer of neutron-proton-electron fluid of about 2~km, and
a thin crust of several 100~m. Clearly, these objects do not resemble
a strange star that is largely composed of quark matter core with a
thin crust floating on top of it, due to support provided by Coulomb
forces~\cite{Weber2005}.

As seen from Figs.~\ref{fig:illust} and~\ref{fig:MR_twi}, all
nucleonic branches are fully compatible with the masses and radii
inferred by the NICER instrument for both parameter sets corresponding
to stiff-soft(stiff) and soft-stiff(soft) nucleonic EoS.

There is no tension between the NICER inferences and the soft EoS
needed to account for GW170817 analysis, as implied by statistical
models~\cite{Dietrich2020Sci,Landry2020PhRvD,Raaijmakers2021ApJ,Biswas2021ApJ}, including those which allow for first-order phase
transitions~\cite{Pang2020PhRvR,Legred2021PhRvD,Pang2021ApJ}. 
There is also no evidence for the need for strong first-order phase 
transition in the analysis of statistical models, while some of them
disfavoring such transition~\cite{Pang2021ApJ}. The large value of
$L_{\rm sym}$ suggested by an analysis~\cite{Reed2021} of the
PREX-II experiment (but see also Ref.~\cite{Reinhard2021})
requires a hard nucleonic EoS which becomes consistent with the
GW170817 event in the case of the H-Q phase
transition~\cite{Lijj2021,Christian2021b}. For an
alternative which used nonparametric EoS and constrains the
symmetry energy and its slope directly from observations with
minimal modeling assumptions, see~Ref.~\cite{Essick2021}.

\begin{table}[b]
\centering
\caption{
Parameters of the used EoS with a fixed maximum mass of the hadronic branch $M^{\rm{H}}_{\rm{max}}= 2.0\,M_\odot $ 
and speed of sound squared $s = 1.0$ in quark matter. The maximum masses of ultracompact hybrid stars are shown in the range $M^{\rm{Q}}_{\rm{max}}/M_\odot = 1.00$-1.60  with a step of 0.20. The remainder columns specify the corresponding values of $\Qsat$, $\Lsym$, and $\varepsilon_{\rm{tran}}$ and 
$\Delta\varepsilon/\varepsilon_{\rm{tran}}$. The last column shows the range of masses within which twin ultracompact stars exist. }
\setlength{\tabcolsep}{9pt}
\label{tab:Parameters2}
\centering
\begin{tabular}{cccccc}
\hline\hline
$Q_{\rm{sat}}$ &$L_{\rm{sym}}$ &$\varepsilon_{\rm{tran}}$ &
$M^{\rm{Q}}_{\rm{max}}$ &$\Delta\varepsilon/\varepsilon_{\rm{tran}}$ &$\Delta M_{\rm{twin}}$ \\
\hline
300  & 45 & 487.132 & 1.00 & 5.0604 & 0.0133 \\ 
     &    &         & 1.20 & 3.3866 & 0.0077 \\
     &    &         & 1.40 & 2.3407 & 0.0065 \\
     &    &         & 1.60 & 1.6325 & 0.0109 \\     
300  &105 & 463.315 & 1.00 & 5.2178 & 0.0283 \\ 
     &    &         & 1.20 & 3.4885 & 0.0232 \\
     &    &         & 1.40 & 2.4100 & 0.0223 \\
     &    &         & 1.60 & 1.6856 & 0.0288 \\ 
900  & 45 & 433.640 & 1.00 & 5.5915 & 0.0411 \\ 
     &    &         & 1.20 & 3.7446 & 0.0350 \\
     &    &         & 1.40 & 2.6015 & 0.0341 \\
     &    &         & 1.60 & 1.8358 & 0.0412 \\   
900  &105 & 414.476 & 1.00 & 5.7823 & 0.0595 \\ 
     &    &         & 1.20 & 3.8663 & 0.0547 \\
     &    &         & 1.40 & 2.6857 & 0.0551 \\
     &    &         & 1.60 & 1.8988 & 0.0635 \\    
\hline\hline
\end{tabular}
\end{table}
The range of masses of such stars can be remarkably broad when varying
the EoS (not the central pressure) and covering the interval
$1.0 \le M/M_{\odot}\le 2.0$. Thus, a consequence of high-density QCD
phase transition is the existence of ultracompact stars --- a
prediction that is consistent with the astrophysical constraints
obtained to date. The parameters fully characterizing the hybrid EoS models that which produce ultracompact stars are given in Table~\ref{tab:Parameters2}.
We note that the only other models that predict such
small radii that may be consistent with the current observational data
are those based on the idea of strange stars~\cite{Weber2005}.

\begin{figure}[tb]
\centering
\ifpdf
\includegraphics[width = 0.45\textwidth]{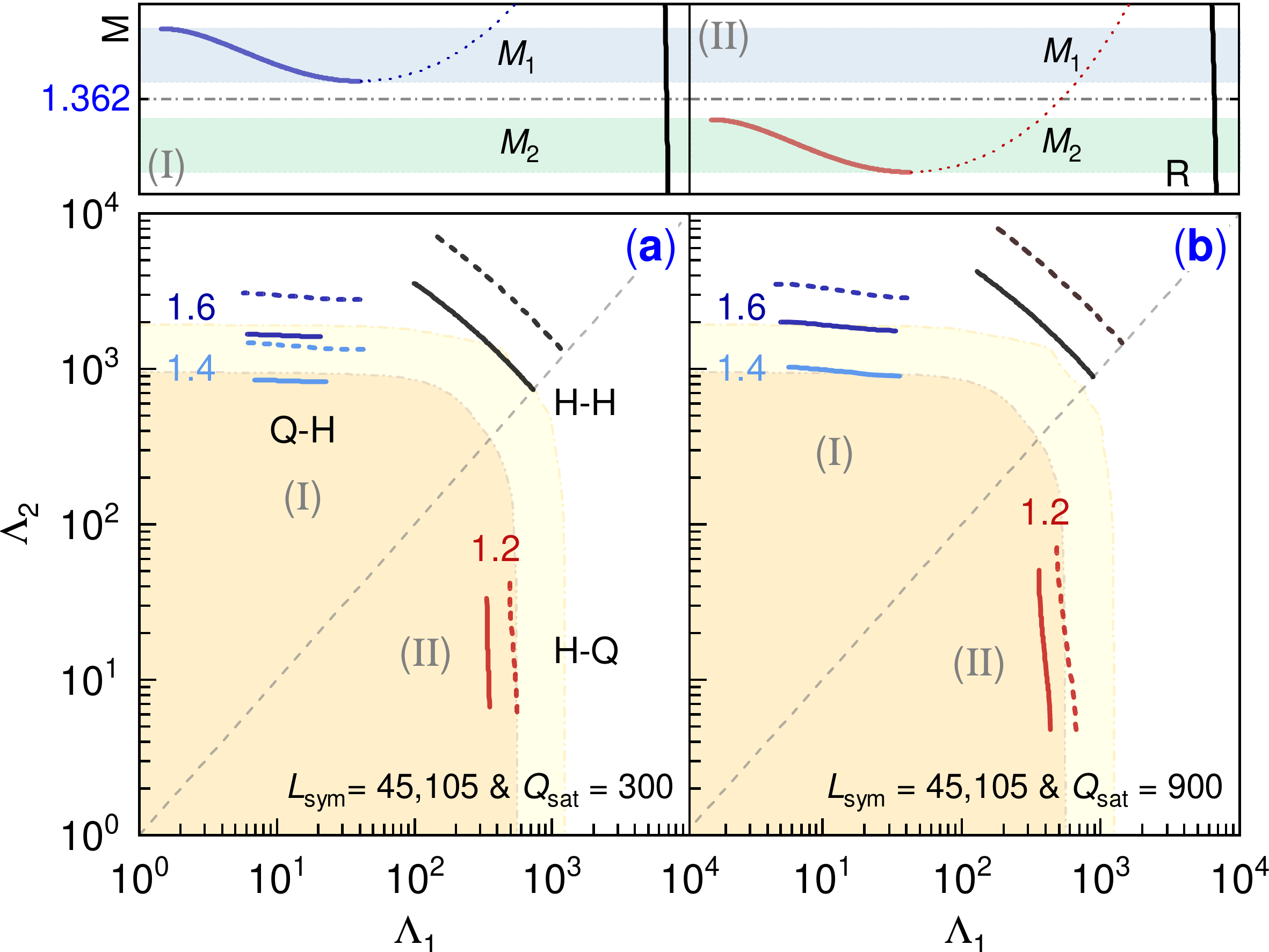}
\else
\includegraphics[width = 0.45\textwidth]{Lam_Lam.eps}
\fi
\caption{ Dimensionless TD of compact objects for a fixed value of
  binary chirp mass $\mathcal{M}/M_\odot= 1.186$. The shaded regions
  correspond to the 50\% and 90\% CIs taken from the
  analysis of GW170817 event.  The EoS models with strong first-order
  phase transition are fixed by the values
  $M^{\rm H}_{\rm max}/M_{\odot}= 2.0$ and
  $M^{\rm Q}_{\rm max}/M_{\odot}= 1.2$ (red curves), 1.4 (blue
  curves), 1.6 (dark blue curves), and $s=1.0$. The type of binary
  components (Q/H) with mass $M_1$-$M_2$ are indicated in the plot.
  Panel (a): soft-soft and stiff-soft hadronic EoS results are shown
  by solid and dashed lines; panel (b): soft-stiff and stiff-stiff
  hadronic EoS are shown by solid and dashed lines. The upper left
  inset shows a schematic $M$-$R$ diagram for the cases
  $M^{\rm Q}_{\rm max}/M_{\odot}= 1.4,\, 1.6$, where the binary is of
  the Q-H type and the hybrid branch lies above the average value
  $M/M_{\odot}= 1.362$. The right inset shows the
  same for $M^{\rm Q}_{\rm max}/M_{\odot}= 1.2$, in which case the
  binary is of the H-Q type with the hybrid branch lying below the
  average value $M/M_{\odot}= 1.362$. These two
  cases are marked in the plots and insets by Latin numerals I and II,
  respectively.  }
\label{fig:GW_lam}
\end{figure}

As demonstrated by multiple analyses of the GW170817 event, the TDs
provide excellent diagnostics of the gross properties of neutron
stars. Note that TDs provide information on the cold EoS of dense
matter, as they are probes originating from the pre-merger phase of
binary inspiral. Figure~\ref{fig:GW_lam} compares our theoretical TDs
for hybrid stars with the observational constraints for this quantity
obtained from the analysis of the GW170817
event~\citep{LIGO_Virgo2018}. Our comparison adopts the chirp mass as
$\mathcal{M} = 1.186\,M_\odot$ and utilizes only the analysis which
assumes the (more plausible) low-spin case~\cite{LIGO_Virgo2018}.  The
masses of the members of the binary in the GW170817 are found to be in
the range $1.16$-$1.60\,{M}_{\odot}$ at 90\% CI.  To
obtain the TDs we (a) fix the values of $\Lsym$ and
$\Qsat$ which selects the nuclear EoS; (b) choose
$M^{\rm H}_{\rm max}/M_{\odot}= 2.0$, 
$M^{\rm Q}_{\rm max}/M_{\odot}= 1.2,$ 1.4, 1.6 and $s=1.0$ which fixes
three hybrid EoS each corresponding to a value of
$M^{\rm Q}_{\rm max}/M_{\odot}$.

Figure~\ref{fig:GW_lam} shows the mutual dependence of TDs
$\Lambda_1$-$\Lambda_2$ of members of a binary for three choices of
$M^{\rm Q}_{\rm max}/M_{\odot}$. The shaded areas correspond to the
50\% and 90\% CIs as indicated in the plot (we adopt the
results obtained from the PhenomPNRT waveform
model~\citep{LIGO_Virgo2019}).  Note that the diagonal on this plot
corresponds to the case of an equal-mass binary with
$M_{1,2} = 1.362\,M_\odot$.  As seen, the case of H-H binary generates
$\Lambda_1$-$\Lambda_2$ values at the boundary or outside of the 90\%
CI region for any choice of the stiffness of the EoS.  In the case of
$M^{\rm Q}_{\rm max}/M_{\odot} =1.4$ and 1.6 the mass range of the
hybrid branch is above the average value $M/M_{\odot} = 1.362$ (as
indicated in the left inset showing \MR diagram). In the case
$M^{\rm Q}_{\rm max}/M_{\odot} =1.2$ the opposite is the case, i.e.,
the hybrid branch is below this value, see the right inset.  As a
consequence, the $\Lambda_1$-$\Lambda_2$ tracks in the first case are
in the upper half of the plot (Q-H binaries). In the second case, they
are in the lower half of the plot (H-Q binaries).  These models are
compatible with the range determined for GW170817 for soft-soft,
stiff-soft (panel a) and soft-stiff, stiff-stiff (panel b) EoS. An
exception is the case $M^{\rm Q}_{\rm max}/M_{\odot} = 1.6$ when the
intermediate-density EoS is chosen to be stiff. From the analysis
above one may conclude that the observations can be explained by 
appropriate choices of nuclear and quark EoS.

\begin{figure}[tb]
\centering
\ifpdf
\includegraphics[width = 0.45\textwidth]{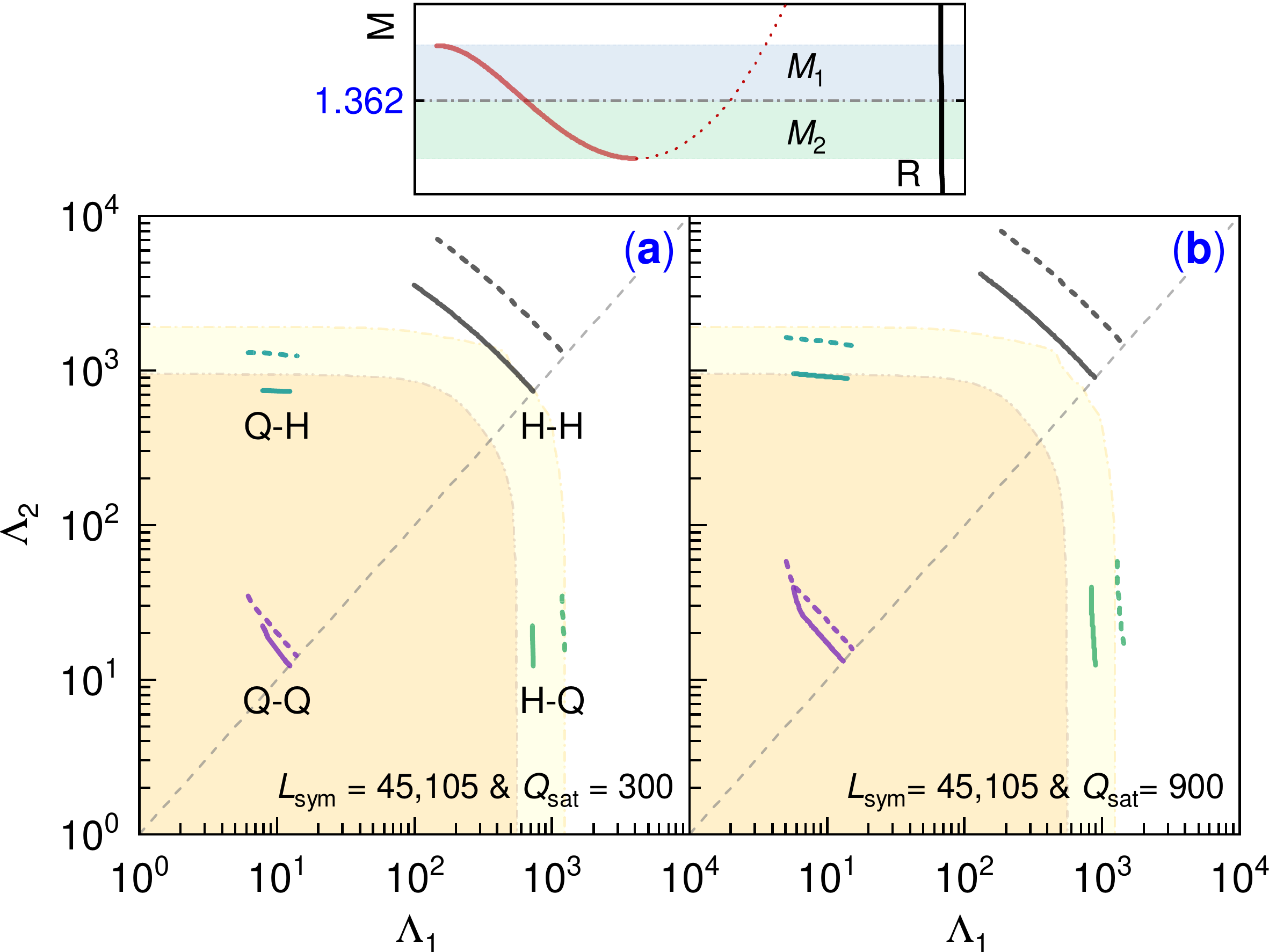}
\else
\includegraphics[width = 0.45\textwidth]{Lam_Lam_2.eps}
\fi
\caption{Dimensionless TD for the case of EoS with the hybrid branch
  covering a range of masses around the average value
  $M/M_{\odot}= 1.362$ as illustrated in the upper
   inset showing the \MR relation. The remaining parameters are
  chosen as in Fig.~\ref{fig:GW_lam}. There are four types of binaries
  composed of hadronic (H) and/or hybrid (Q) stars. Panel (a):
  soft-soft and stiff-soft hadronic EoS results are shown with solid
  and dashed lines; (b): soft-stiff and stiff-stiff hadronic EoS are
  shown by solid and dashed lines.}
\label{fig:GW_lam2}
\end{figure}

Figure~\ref{fig:GW_lam2} shows the same as in Fig.~\ref{fig:GW_lam}, but
in this case, where the quark matter EoS is chosen so that the hybrid branch
covers a range of masses whose average value is $1.362\,M_\odot$,
see the inset. Each member of the binary can be chosen from either of 
H or Q branches, so four combinations of stars are possible. It is 
seen that all the pairs lie within the allowed range except the 
H-H pair which is at the boundary or outside of 90\% CI.  Interestingly, 
the nucleonic EoS of hybrid stars need not be soft. Even with an EoS 
that is stiff at intermediate densities (stiff-stiff or stiff-soft EoS), 
a Q-Q binary, consisting of two ultracompact hybrid stars with very 
small radii can have a combined TD as low as $\tilde\Lambda_{1.186} \simeq 10$.  
This suggests that the multimessenger data can be accounted for if PSR
J0740+6620 and J0030+0451 are purely nucleonic stars, whereas the two
components of GW170817 binary are ultracompact hybrid stars.

\section{Conclusions}
To summarize, we have uncovered several implications of the strongly 
first-order phase transitions at high density in the QCD phase diagram 
that are of fundamental importance for the analysis of the data from 
current and future gravitational-wave observatories, X-ray missions 
and terrestrial experiments aimed at the determination of the skin of
nuclei. Firstly, we have confirmed that the nuclear EoS does not need
to be soft to account for the GW170817 event, as the phase transitions
allow for the emergence of hybrid stars with properties consistent
with the GW170817 analysis \cite{Lijj2021,Christian2021b}.  The
NICER inferences are accounted for by invoking pure nucleonic stars
that live on the hadronic (second family) branch of compact stars and
consistency with an analysis of the PREX-II is achieved (which was
precluded in models without phase transition). Secondly, we determine
the range of the masses for which twin configurations, i.e.,
identical-mass stars with different radii, arise. This analysis shows
that the priors in the statistical analysis of the data should be
constructed consistent with the possibility of a first-order phase
transition with star masses in a wide range
$1.0\le M/M_{\odot}\le 2.5$ and radii covering {\it also the range of
small radii 6-9 km which was thought to be only achievable for strange
stars, and is currently excluded by the NICER analysis at 90\% CI.}
 
In our proposal, both branches of stars consist of stars with thick
nucleonic crusts, consistent with observations of surface phenomena
such as seismic vibrations after giant flares in magnetars
\cite{Watts2006,Chugunov2006,Suvorov2022}, the thermal response of the
crust to accretion \cite{Brown2000,Potekhin2021}, the contribution of
the crust to the moment of inertia and glitches
\cite{Glendenning1992,Haskell2018}, etc.  This provides an alternative
to scenarios such as that proposed in Ref.~\cite{Drago2013} where one
of the branches consists of strange/quark stars with a very thin
nucleonic crust.

\section*{Acknowledgments}  
M.~A. is partly supported by the U.S. Department of Energy, Office of
Science, Office of Nuclear Physics under Award No. DE-FG02-05ER41375.
J.~L. is supported by the National Natural Science Foundation of China
(Grant No. 12105232), the Fundamental Research Funds for the Central
Universities (Grant No. SWU-020021), and by the Venture \& Innovation
Support Program for Chongqing Overseas Returnees (Grant
No. CX2021007).  The research of A.~S. was funded by Deutsche
Forschungsgemeinschaft Grant No. SE 1836/5-2 and the Polish NCN Grant
No. 2020/37/B/ST9/01937 at Wroc\l{}aw University.

%

\end{document}